\documentclass[prl,twocolumn,showpacs,floatfix,amsfonts]{revtex4}
\usepackage{graphicx,graphics,color,epsfig}
\usepackage{bm}
\usepackage{amsmath}
\usepackage{amssymb}
\usepackage{fancyheadings}
\begin{document}
\preprint{}
\title{Vibration Mode Induced Shapiro Steps and Backaction
in Josephson Junctions}
\author{Jian-Xin Zhu, Z. Nussinov, and A. V. Balatsky}
\affiliation{Theoretical Division, Los Alamos National Laboratory,
Los Alamos, New Mexico 87545}

\begin{abstract}
A model of a superconducting tunnel junction coupled to a
mechanical oscillator is studied at zero temperature in the case
of linear coupling between the oscillator and tunneling electrons.
It is found that the Josephson current flowing between two
superconductors is modulated by the motion of the oscillator.
Coupling to harmonic oscillator produces additional Shapiro steps
in $I-V$ characteristic of Josephson junction whose position is
tuned by the frequency of the vibration mode. We also find a new
velocity dependent term originating from the backaction of the
ac Josephson current. This term is periodic in time and 
vanishes at zero bias voltage.
\end{abstract}
 \maketitle

The coupling of the charge carriers to vibrational modes and localized
spins in electronic
devices has been a subject of intense investigation recently. Vibrational
modes and spins possess dynamic internal degrees of freedom,
much unlike static impurities or defects.
As a consequence, they affect the electronic dynamics in these devices.
Interesting $I$-$V$ characteristics (i.e., peaks in differential
tunneling conductance) in molecular
electronics~\cite{multi1,Aviram98,Langlais99,Park00,Park02,Zhitenev02}
may indicate strong influence from the electronic-vibrational
coupling. A step structure (rather than peak structure) in the
differential tunneling conductance has also been observed in the
STM-based inelastic tunneling spectroscopy around a local
vibrational mode on surfaces~\cite{Stipe99}. The vibrational
effects on the conductance of molecular quantum dots were also
examined ~\cite{Wingreen89,Ludin02,Zhu03a,Mitra03,Flensberg03,Aji03}.

Spin detection and manipulation is crucial in spintronics and
quantum information processing. The modulation of the current by
precessing spins may be used to detect and manipulate single
spins in electron-spin-resonance-scanning tunneling microscopy
(ESR-STM) technique~\cite{Mana89,Mana00,Durkan02,Manoharan02}.
Theoretically, it is
known~\cite{Balatsky02a,Balatsky02b,Zhu02,Mozy01} that
spin-orbit interactions and the polarization of injected electrons
are important in modulating the tunneling current. Theses studies
focused on the tunneling between two normal metals, where
only the single particle process is involved in transport.
Recently, we studied a precessing spin coupled through a
local direct exchange interaction to the tunneling electrons
between two superconductors~\cite{Zhu03b}.

We now consider, for the first
time, the Josephson effect in a superconducting tunneling junction
coupled to a mechanical oscillator in the tunnel barrier. Both
effects of the oscillator motion on the tunnel current and the
tunneling electrons on the oscillator dynamics are studied. To our
knowledge, none of these effects have been addressed before.
We find that: (i) In the tunnel junction, the
Josephson current is modulated by the motion of the oscillator;
the Fourier spectrum of the Josephson current exhibits peaks at
frequency $\omega_{J}\pm \omega_{0}$, $2\omega_{J}$, and
$\omega_{J} \pm 2\omega_{0}$, where $\omega_{J}=2eV$ and
$\omega_{0}$ are the Josephson frequency and the vibrational mode
of the oscillator respectively. These additional peaks are the
result of coupling of ac Josephson current at $\omega_J$ and
oscillation at $\omega_0$.  (ii) Electron tunneling
through the junction leads to a novel
time dependent change in oscillator energy.
If no voltage bias is applied across the junction
(i.e., dc Josephson effect in equilibrium), the oscillator energy is
time independent. When a
nonzero voltage bias applied, the oscillator will display
time-dependent energy variations.

\begin{figure}
\includegraphics*[width=\columnwidth]{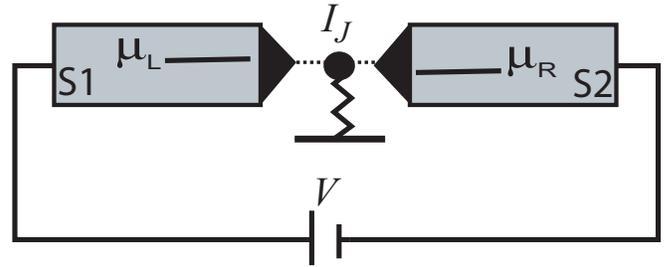}
\caption{An oscillator in the tunneling barrier coupled to two
superconducting leads. The motion of the oscillator modulates the
electron tunneling between the superconductors. The chemical
potentials on both sides differ by the applied voltage,
$\mu_{L}-\mu_{R}=eV$. } \label{FIG:SETUP}
\end{figure}

Our model is illustrated in
Fig.~\ref{FIG:SETUP}. It consists of two ideal superconducting
leads coupled to each other by a tunneling barrier, which contains
a mechanical oscillator. The Hamiltonian
\begin{equation}
H=H_{L}+H_{R}+H_{T}\;. \label{EQ:Hamil}
\end{equation}
The first two terms are, respectively, the Hamiltonians for electrons
in the left and right superconducting leads of the tunnel
junction,
\begin{eqnarray}
&H_{L(R)}=\sum_{\mathbf{k}(\mathbf{p});\sigma}\epsilon_{\mathbf{k}(\mathbf{p})}
c_{\mathbf{k}(\mathbf{p}),\sigma}^{\dagger}c_{\mathbf{k}(\mathbf{p}),\sigma}&
 \nonumber \\
& +\sum_{\mathbf{k}(\mathbf{p})} [\Delta_{L(R)}
c_{\mathbf{k}(\mathbf{p})\uparrow}^{\dagger}
c_{-\mathbf{k}(-\mathbf{p})\downarrow}^{\dagger} +\mbox{H.c.}]\;,&
\end{eqnarray}
where we denote the electron creation (annihilation)
operators in the left,right (L,R) leads by $c_{\mathbf{k}\sigma}^{\dagger}$
($c_{\mathbf{k}\sigma}$) and
$c_{\mathbf{p}\sigma}^{\dagger}$ ($c_{\mathbf{p}\sigma}$) respectively.
$\mathbf{k}$ ($\mathbf{p})$ are momenta and $\sigma$ is
the spin index, $\epsilon_{\mathbf{k}(\mathbf{p}),\sigma}$ and
$\Delta_{L(R)}$ are, respectively, the single particle energies of
conduction electrons, and the pair potential (the gap
function) in the leads. Without loss of generality, we assume that
the superconductors are of a spin-singlet $s$-wave pairing
symmetry, and consider the Josephson tunneling at zero
temperature.
The third term in Eq.~(\ref{EQ:Hamil}) depicts the tunneling
between the superconductors:
\begin{equation}
H_{T}=\sum_{\mathbf{k},\mathbf{p};\sigma} [T_{\mathbf{kp}}
c_{\mathbf{k}\sigma}^{\dagger}c_{\mathbf{p}\sigma}
+\mbox{H.c.}]\;,
\end{equation}
where the tunneling matrix elements $T_{\mathbf{kp}}$ transfer
electrons through an insulating barrier. When a local vibrational
mode is embedded into the tunneling barrier then, in the linear
coupling regime,
\begin{equation}
T_{\mathbf{kp}}=T_{\mathbf{kp}}^{(0)}[1+\alpha u],
\end{equation}
where $\alpha$ describes the coupling between the tunneling
electrons and vibrational mode. The quantity $u$ is the
displacement operator for the oscillator.

Henceforth, we denote by $M$ and $K$
the mass and spring constant of
the mechanical oscillator. As
the energy associated with the vibrational mode,
$\omega_{0}=\sqrt{K/M}\sim 10^{-1} - 10^{-6}\;\mbox{eV}$ is much smaller
than the typical electronic energy on the order of 1 eV, the
mechanical oscillation is very slow as compared to the time scale
of electronic processes. This allows us to apply the
{\em Born-Oppenheimer approximation} to treat the electronic
degrees of freedom as if the local oscillator is static at every
instantaneous location. We will treat the dynamics of the
mechanical oscillator including the back action from the tunneling
electrons.

Whenever a voltage bias is applied across the junction, the Josephson
current
\begin{eqnarray}
I_{J}(t)&=&e\int_{-\infty}^{t} dt^{\prime} [e^{ieV(t+t^{\prime})}
\langle [A(t),A(t^{\prime})]_{-}\rangle \nonumber \\
&& -e^{-ieV(t+t^{\prime})} \langle
[A^{\dagger}(t),A^{\dagger}(t^{\prime})]_{-}\rangle]\;,
\label{EQ:JOSEPHSON}
\end{eqnarray}
where the operator $A(t)=\sum_{\mathbf{k},\mathbf{p};\sigma}
T_{\mathbf{kp}} \tilde{c}_{\mathbf{k}\sigma}^{\dagger}(t)
\tilde{c}_{\mathbf{p}\sigma}(t)$. Here
$\tilde{c}_{\mathbf{k}(\mathbf{p})\sigma}(t) =e^{iK_{L(R)}t}
c_{\mathbf{k}(\mathbf{p})\sigma}e^{-iK_{L(R)}t}$ with
$K_{L(R)}=H_{L(R)}-\mu_{L(R)}N_{L(R)}$ and
$N_{L(R)}=\sum_{\mathbf{k}(\mathbf{p}),\sigma}
c_{\mathbf{k}(\mathbf{p})\sigma}^{\dagger}
c_{\mathbf{k}(\mathbf{p})\sigma}$. The unequal chemical potentials
of the two superconductors lead to a voltage bias
$\mu_{L}-\mu_{R}=eV$. Hereafter, we set $\hbar=1$. A little algebra
yields:
\begin{eqnarray}
I_{J}(t)&=&J_{S}^{(0)}(eV) [1+\alpha u]^{2} \sin (\omega_{J}t)
\nonumber \\
&&+\Gamma_{S}(eV)(1+\alpha u) \alpha \frac{\partial u}{\partial t}
\cos(\omega_{J}t) \;. \label{EQ:Josephson-0}
\end{eqnarray}
Here we assume that the two superconductors are identical and set
the constant phase difference between them $\phi_{0}=0$. The
Josephson frequency is given by $\omega_{J}=2eV$. The quantity
$J_{S}^{0}$ is the amplitude of the Josephson current in the
absence of coupling to the vibrational mode (that is, $\alpha=0$),
which is found to be,
\begin{eqnarray}\label{EQ:JS2}
J_{S}^{(0)}(eV)&=&e\sum_{\mathbf{k},\mathbf{p}}\frac{\vert
\Delta\vert^{2}\vert
T^{(0)}_{\mathbf{kp}}\vert^{2}}{E_{\mathbf{k}}E_{\mathbf{p}}}\biggl{(}
\frac{1}{eV+E_{\mathbf{k}}+E_{\mathbf{p}}} \nonumber \\
&&-\frac{1}{eV-E_{\mathbf{k}}-E_{\mathbf{p}}}\biggr{)}\;,
\end{eqnarray}
with $\vert\Delta\vert$ is the superconducting energy gap and
$E_{\mathbf{k}(\mathbf{p})}=\sqrt{(\epsilon_{\mathbf{k}(\mathbf{p})}-E_{F})^{2}
+ \vert
\Delta \vert^{2}}$. The amplitude
\begin{eqnarray}
\Gamma_{S}(eV)&=&e\sum_{\mathbf{k},\mathbf{p}}\frac{\vert
\Delta\vert^{2}\vert
T^{(0)}_{\mathbf{kp}}\vert^{2}}{E_{\mathbf{k}}E_{\mathbf{p}}}\biggl{[}
\frac{1}{(E_{\mathbf{k}}+E_{\mathbf{p}}-eV)^{2}}\nonumber \\
&&-\frac{1}{(E_{\mathbf{k}}+E_{\mathbf{p}}+eV)^{2}}\biggr{]}\;.
\label{EQ:GammaS}
\end{eqnarray}
An order of magnitude estimate gives $\Gamma_S \omega_{J}/
J_{S}^{(0)} \sim (eV/\vert \Delta \vert)^{2}\ll 1$.

From Eq.~(\ref{EQ:Josephson-0}), we may construct the coupling
modulated part of the Josephson junction energy
\begin{eqnarray}\label{EQ:HJ}
H_{J}&=&E_{J} (1+\alpha u)^{2} [1-\cos(\omega_{J}t)] \nonumber
\\
&& +\frac{\Gamma_{S}}{2e} \alpha (1+\alpha u) \frac{\partial
u}{\partial t} \sin(\omega_{J}t)\;,
\end{eqnarray}
where $E_{J}=J_{S}^{(0)}/2e$. The derivative of $H_J$ with respect to
the phase yields the supercurrent in
Eq.(\ref{EQ:Josephson-0}). Eq.(\ref{EQ:HJ}) captures the
{\em Josephson backaction effect}- it vividly illustrates
how {\em the electronic degrees
of freedom influence the mechanical oscillator} of Fig.1 coupled
to them. We may obtain this
result by also employing other standard techniques.
For instance, integrating out the fermionic degrees of freedom,
we obtain an effective action $S_{eff}$
for $u(\tau)$ in Matsubara time.
When the standard mechanical part of the oscillator is
added, the action
\begin{eqnarray}
S&=& \frac{1}{2} \int d \tau \int d \tau^{\prime} ~
|T^{(0)}_{\bf{kp}}|^{2} (1+ \alpha u(\tau)) (1+ \alpha
u(\tau^{\prime})) K(\tau,\tau^{\prime}) \nonumber
\\
&&+ \frac{1}{2} \int d \tau [K u^{2} + M (\frac{d u}{d
\tau})^{2}]. \label{EQ:Matsubara}
\end{eqnarray}
The non-locality in time present in the first
term reflects how electronic correlations
between different times influence the oscillator
coordinates and lead to the Josephson backaction effect.
Here, $K(\tau,\tau^{\prime}) = F_{L}(\tau-\tau^{\prime})
F_{R}(\tau^{\prime} - \tau) \exp[eV(\tau + \tau^{\prime})]$,
with $F$ the Matsubara-Gorkov function. As
the resulting classical action is not time translationally
invariant, energy is not conserved. Within the Born-Oppenheimer approximation
($u(\tau^{\prime}) \simeq u(\tau) + (\tau^{\prime} - \tau)
\partial u/\partial \tau$), the equation of motion
\begin{eqnarray}
0= \frac{\delta S}{\delta u(\tau)} = M \frac{d^{2}u}{dt^{2}}
+ K u \nonumber
\\
+
\alpha |T_{\bf{kp}}^{(0)}|^{2} \int d\tau^{\prime}
K(\tau,\tau^{\prime}) [1+ \alpha u(\tau) + \alpha \frac{\partial u}{\partial \tau}
(\tau^{\prime} - \tau)],
\end{eqnarray}
leading to
\begin{equation}
M\frac{d^{2}u}{dt^{2}}+\gamma_{S}(t) \frac{\partial u}{\partial t}
+Ku
=F(t)\;. \label{EQ:Oscillator}
\end{equation}
Here, the driving force
\begin{equation}
F(t)=-2\alpha E_{J} (1+\alpha u)\biggl{\{}
1-[1+\frac{\Gamma_{s}\omega_{J}}{4eE_{J}}]\cos(\omega_{J}t)\biggr{\}}\;,
\label{EQ:Force}
\end{equation}
with $\Gamma_{S}\omega_{J}/4eE_{J} \simeq (eV/\vert \Delta
\vert)^{2} \ll 1$, and the time-dependent energy non-conserving
\begin{equation}
\gamma_{S}(t)=-(\alpha^{2} \Gamma_{s}/e)\sin(\omega_{J}t)\;.
\label{EQ:Damping}
\end{equation}
These relations are also very transparent within the real time
approach of Eq.~(\ref{EQ:HJ}), where the total oscillator
Hamiltonian $H_{osc} = H_{J}+ P^{2}/2M+ Ku^{2}/2$. A more detailed
Keldysh contour analysis yields the same results
\cite{us-Keldysh}. In Eq.(\ref{EQ:Force}), the first, linear in
$u$ term, may, alternatively, be lumped into the spring moduli $K$
and regarded as a {\em Josephson stiffness}- a shift of the spring
constant resulting from electronic correlations. The oscillatory
part of the driving force and the time dependence of the Josepshon
backaction generated $\gamma_{S}(t)$ (Eq.(\ref{EQ:Damping})) lead
to interesting experimental consequences which we will now
elaborate on. 
Equations~(\ref{EQ:Josephson-0}), (\ref{EQ:Oscillator}),
(\ref{EQ:Force}) and (\ref{EQ:Damping}) constitute the central
result of this work.

{\em Josephson Bakaction.} As shown by Eq.~(\ref{EQ:Damping}) and
Eq.~(\ref{EQ:GammaS}), the velocity coefficient ($\gamma_{S}(t)$) originates
from the coupling of the mechanical oscillator to the tunneling electrons.
It is quadratically proportional to the coupling constant
$\alpha$, which is similar to the case of normal metal tunnel
junctions~\cite{Mozy02b}. This term has two novel features:
(i) $\gamma_{S}$ depends on the
voltage bias. At zero voltage bias (the dc setting),
$\gamma_{S}$ vanishes since
$\Gamma_{S}$ is zero. $\gamma_{S}$ is finite
only when a finite voltage bias is applied
across the junction (i.e., ac case). In the low voltage limit ($eV
\ll \vert \Delta \vert$), $\gamma_{S}$ is linearly proportional to
the voltage bias. (ii) Once a finite voltage bias is applied,
$\gamma_{S}$ is also a periodic function of time with the
Josephson frequency $\omega_{J}$. Both of these features are
absent in the normal metal tunnel junctions~\cite{Mozy02b}.
These properties
are unique to the coupling of the mechanical oscillator to the
superconductors. In the normal metal, there is no quasiparticle
energy gap on the Fermi surface. Matters become far richer when the
oscillator is coupled to the superconductor. On the one hand,
in the superconductor, there
exists an energy gap on the Fermi surface and the quasiparticles
are depleted below this energy. This leads to the quenching of the
single particle tunneling channel; no contribution to the
dissipation of the oscillator due to normal quasiparticles is
possible. On the other hand, due to macroscopic quantum coherence
in the superconductor, Cooper pairs can tunnel through the barrier
between the two superconductors with a probability
comparable to that of single particle tunneling in a normal metal
junction.
When a static voltage bias is applied, the tunneling of a single
Cooper pair requires an energy $2eV$
to overcome the potential barrier. The energy carried by Cooper
pair upon tunneling can be transferred to oscillator. 
For zero bias voltage, the tunneling
pairs do not acquire/lose
energy (the effective electronic only action is time
independent, no effective external sources
are present, and energy is preserved
at all times). In the ac setting,  $\sin
\omega_J t$ is odd under time reversal and, as a
consequence, the term $\sin
\omega_J t \frac{\partial u}{\partial t}$ is allowed in
the effective Hamiltonian Eq.~(\ref{EQ:HJ}).

\begin{figure}
\includegraphics*[width=8cm,height=8cm]{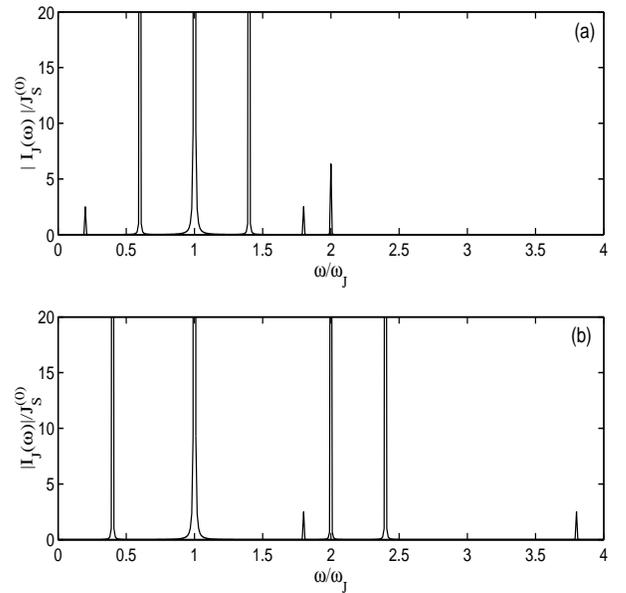}
\caption{Absolute magnitude of the Fourier transform
$I_{J}(\omega)$ of the Josephson current as given by
Eq.~(\ref{EQ:Josephson-1}) for the vibration mode $\omega_{0}=0.4
\omega_{J}$ (a) and $1.4$ (b). The other parameters values are
taken to be: $\tilde{\alpha}=0.1$, and $\tilde{K}=0.6$. }
\label{FIG:Fourier}
\end{figure}

{\em Shapiro steps.} To calculate the Josephson current, we need
to solve Eq.~(\ref{EQ:Oscillator}) for the displacement field
$u(t)$. In the weak coupling limit and in view of the fact that
$\Gamma_{S}$ is much smaller than $J_{S}^{(0)}$, the main physics
can be captured by neglecting the damping terms and the $\alpha
\Gamma_{S}$ and $\alpha^{2}$ terms in the driving force. In this
limit,
\begin{equation}
u(t)=u_{0}\cos(\omega_{0}t)+\frac{2\alpha E_{J}}{M(\omega_{0}^{2}
-\omega_{J}^{2})}\cos(\omega_{J}t)-\frac{2\alpha E_{J}}{K}\;,
\label{EQ:Displacement}
\end{equation}
and the Josephson current
\begin{eqnarray}
I_{J}(t)&=&J_{S}^{(0)}\biggl{\{}
(1-\frac{4\tilde{\alpha}^{2}}{\tilde{K}})\sin(\omega_{J}t)
 +2\tilde{\alpha}\cos(\omega_{0}t)\sin(\omega_{J}t)
\nonumber
\\
&&+\frac{2\tilde{\alpha}^{2}\sin(2\omega_{J}t)}{\tilde{K}
(1-\omega_{J}^{2}/\omega_{0}^{2})}
 +\tilde{\alpha}^{2}\cos^{2}(\omega_{0}t)
\sin(\omega_{J}t)\biggr{\}} \;, \label{EQ:Josephson-1}
\end{eqnarray}
where $\tilde{\alpha}=\alpha u_{0}$ and
$\tilde{K}=Ku_{0}^{2}/E_{J}$. Equation~(\ref{EQ:Josephson-1})
demonstrates clearly that the Josephson current not only
oscillates with time with a frequency $\omega_{J}$, but is also
modulated by the vibrational mode of the mechanical oscillator
with a frequency $\omega_{0}$. In Fig.~\ref{FIG:Fourier}, we plot
the absolute magnitude of the Fourier transform of the Josephson
current given by Eq.~(\ref{EQ:Josephson-1}) for various values of
the vibration mode frequency $\omega_{0}$. The spectrum shows a
main peak at the frequency $\omega_{J}$. In addition, the coupling
of the mechanical oscillator and tunneling electrons generates new
side peaks at frequencies $\omega_{J}\pm \omega_{0}$,
$\omega_{J}\pm 2\omega_{0}$, and $2\omega_{J}$. The intensity of
these peaks is proportional to the coupling constant. Note that
the peak at $2\omega_{J}$ originates from the second term in
$u(t)$ given by Eq.~(\ref{EQ:Displacement}), which is a direct
manifestation of the feedback effect from the Josephson tunneling.
Our calculation implies that a dc component arises if
the voltage bias has one of the Shapiro step values $\omega_{0}$
and $2\omega_{0}$. When higher-order
effects are taken into account, the equation-of-motion for the
oscillator can only be solved numerically. The main
conclusions presented here remain qualitatively
unchanged.

{\em Conclusion.} We studied the Josephson junction coupled to a
mechanical oscillator between its two superconducting leads. We
found that the Josephson current flowing between two spin-singlet
pairing superconductors is modulated by the motion of the
oscillator. The coupling of an oscillator of eigenfrequency
$\omega_0$ to an ac junction of characteristic frequency
$\omega_J$ leads to beats. We find novel Shapiro steps induced at
$\omega_{J}\pm \omega_{0}$, $\omega_{J}\pm 2\omega_{0}$, and
$2\omega_{J}$. This differs from the case of a precessing
spin~\cite{Zhu03b}, where due to the sum rule of tunneling through
different spin channels, the Josephson current is not modulated if
the spin-orbit coupling mechanism does not exist. The electrons
tunneling through the superconducting junction lead to a novel
non-energy conserving effect. If a voltage bias is applied, this
time dependent effect arises from the back-action of the
supercurrent on the oscillator dynamics. As far as we know, no
measurements of Josephson current through a vibrational mode
between two superconductors have been reported yet. Recent
progress in molecular electronics~\cite{Park00} and nanomechanical
resonators~\cite{Huang03,Blencowe00} holds great promise in
attaching single molecules to superconducting leads, or even tune
the tunnel barrier of the superconducting junctions by a
mechanical cantilever. Our predictions are, potentially, within
the realm of present technology. Another possible experiment
concerns atomically sharp superconducting tip in low temperature
STM in both the quasiparticle tunneling ~\cite{Pan98} and
Josephson tunneling regimes~\cite{Naaman01} (coined ``Josephson
STM'' or JSTM~\cite{Smakov01}) on conventional superconductors. It
is very interesting to extend the JSTM technology by using a
superconducting tip to study the Josephson current in the vicinity
of a local vibrational mode on the superconducting surface, which
may provide a new detection technique.

{\bf Acknowledgments}: We gratefully acknowledge
useful discussions with A. Shnirman.
This work was supported by the U.S.
Department of Energy.


\begin{thebibliography}{99}

\bibitem{multi1} R. P. Andres {\em et al.},
Science {\bf 272}, 1323 (1996); M. A. Reed {\em et al.},
{\em ibid.} 
{\bf 278}, 252 (1997); C. Kergueris {\em et al.},
Phys. Rev. B {\bf 59}, 12505 (1999); J. Reichert {\em et al.},
Phys. Rev. Lett. {\bf 88}, 176804 (2002); S. Hong {\em et al.},
Superlatt. and Microstruc. {\bf 28}, 289 (2000); J. J. W. M.
Rosink {\em et al.},
Phys. Rev. B {\bf 62}, 10459 (2000); J. Chen {\em et al.},
Appl. Phys. Lett. {\bf 77}, 1224 (2000); D. Porath {\em et al.},
Nature {\bf 403}, 635 (2000); R. H. M. Smit {\em et al.},
Nature {\bf 419}, 906 (2002).

\bibitem{Aviram98} A. Aviram and M. Ratner, eds., {\em Molecular
Electronics: Science and Technology} (Annals of the New York
Academy of Sciences, New York, 1998).

\bibitem{Langlais99} V. Langlais {\em et al.},
Phys. Rev. Lett. {\bf 83}, 2809 (1999).

\bibitem{Park00} H. Park {\em et al.},
Nature {\bf 407}, 57 (2000).

\bibitem{Park02} J. Park {\em et al.}, Nature {\bf 417}, 722
(2002).

\bibitem{Zhitenev02} N. B. Zhitenev, H. Meng, and Z. Bao,
Phys. Rev. Lett. {\bf 88}, 226801 (2002).

\bibitem{Stipe99} B. C. Stipe, M. A. Rezaei, and W. Ho, Science
{\bf 280}, 1732 (1998); Phys. Rev. Lett. {\bf 82}, 1724 (1999); N.
Lorente, M. Persson, L. J. Lauhon, and W. Ho, {\em ibid.} {\bf
86}, 2593 (2001).

\bibitem{Wingreen89} N. S. Wingreen, K. W. Jacobsen, and J. W.
Wilkins, Phys. Rev. B {\bf 40}, 11834 (1989).

\bibitem{Ludin02} U. Lundin and R. H. McKenzie, Phys. Rev. B {\bf
66}, 075303 (2002).

\bibitem{Zhu03a} J.-X. Zhu and A. V. Balatsky, Phys. Rev. B {\bf
67}, 165326 (2003).

\bibitem{Mitra03} A. Mitra, I. Aleiner, and A. J. Millis,
cond-mat/0302132.

\bibitem{Flensberg03} K. Flensberg, cond-mat/0302193.

\bibitem{Aji03} V. Aji, J. E. Moore, and C. M. Varma,
cond-mat/0302222.



\bibitem{Mana89} Y. Manassen {\em et al.},
Phys. Rev. Lett. {\bf 62},
2531 (1989); D. Shachal and Y. Manassen, Phys. Rev. B {\bf 46},
4795 (1992); Y. Manassen, J. Magnetic Reson. {\bf 126}, 133
(1997).

\bibitem{Mana00} Y. Manassen, I. Mukhopadhyay, and N. Ramesh Rao,
Phys. Rev. B {\bf 61}, 16223 (2000).

\bibitem{Durkan02} C. Durkan and M. E.
Welland, Appl. Phys. Lett. {\bf 80}, 458 (2002).

\bibitem{Manoharan02} H. Manoharan, Nature {\bf 416}, 24 (2002).

\bibitem{Balatsky02a} A. V. Balatsky and I. Martin, cond-mat/0112407.





\bibitem{Balatsky02b}   A.V. Balatsky, Y. Manassen and R. Salem, Phys.
Rev {\bf B 66}, 195416, (2002).


\bibitem{Zhu02} J.-X. Zhu and A. V. Balatsky, Phys. Rev. Lett.
{\bf 89}, 286802 (2002).

\bibitem{Mozy01} D. Mozyrsky,
L. Fedichkin, S. A. Gurvitz, and G. P. Berman, Phys. Rev. B {\bf
66}, 161313 (2002).




\bibitem{Zhu03b} J.-X. Zhu and A. V. Balatsky, Phys. Rev. B {\bf
67}, 174505 (2003).

\bibitem{us-Keldysh} J-X. Zhu, Z. Nussinov, and A. V. Balatsky,
to be published.

\bibitem{Mozy02b} D. Mozyrsky and I. Martin, Phys. Rev. Lett. {\bf
89}, 018301 (2002).

\bibitem{Huang03} X. M. H. Huang {\em et al.}, Nature {\bf 421},
496 (2003); I. Bargatin and M. L. Roukes, cond-mat/0304605.



\bibitem{Blencowe00} M. P. Blencowe and M. N. Wybourne, Appl.
Phys. Lett. {\bf 77}, 3845 (2000).




\bibitem{Pan98} S. H. Pan, E. W. Hudson, and J. C. Davis, Appl.
Phys. Lett. {\bf 73}, 2992 (2001).

\bibitem{Naaman01} O. Naaman, W. Teizer, and R. C. Dynes, Phys.
Rev. Lett. {\bf 87}, 097004 (2001).

\bibitem{Smakov01} J. \v{S}makov, I. Martin, and A. V. Balatsky, Phys.
Rev. B {\bf 64}, 212506 (2001).

\end{thebibliography}
\end{document}